\documentclass[prd,epsfig,aps,amssymb,amsmath,showpacs]{revtex4}   

\usepackage{graphicx}
\usepackage{amssymb}
\usepackage{amsmath}
\usepackage{color}

\begin{document}

\title{Quantum Chaos Versus Classical Chaos: Why is Quantum Chaos Weaker?}

\author{H. Kr\"{o}ger$\footnote{Corresponding author, Email: hkroger@phy.ulaval.ca}$, J.F. Laprise, G. Melkonyan, R. Zomorrodi}

\affiliation{
{\small\sl D\'{e}partement de Physique, Universit\'{e} Laval, 
Qu\'{e}bec, Qu\'{e}bec G1K 7P4, Canada} 
}

\begin{abstract}
We discuss the questions: How to compare quantitatively classical chaos with quantum chaos? Which one is stronger? What are the underlying physical reasons?
\end{abstract}


\date{\today}

\maketitle

\section{Introduction}
\label{sec:Intro}

Chaotic systems occur in nature everywhere. Examples are: the turbulent flow of gases and liquids, e.g. when pouring milk into a cup of coffee, in the bath tub, in hurricanes, or in the formation of galaxies. Chaos occurs in biology in the dynamics of populations and species. It occurs in the rate of heart beats, 
in attacks of epilepsy, and in the brain. It occurs also in chemical reactions. 
Most of the dynamics occuring in nature is chaotic. The occurence of non-chaotic systems in nature is more of an exception.   
The mathematical description of chaos has reached a mature state over the last decade. The tools being used are classical phase space, Lyapunov exponents, 
Poincar\'e sections, Kolmogorov-Sinai entropy and others.

There is also chaos in the quantum systems 
\cite{Friedrich89,Bohigas93,Guhr98}. People have investigated in quantum physics the analogues of classically chaotic systems. For example, a stadium-shaped billiard is classically chaotic. The corresponding quantum system is a system of ultra-cold atoms bouncing against walls of stadium shape, being created by interaction of the atoms with laser beams. Quantum chaos has been found to play an important role in dynamical tunneling 
\cite{Hensinger,Raizen}. The fingerprints of chaos in quantum systems and the mathematical tools of its description are quite different from those used in classical chaos. The neccessity for a different treatment is due to the nature of quantum mechanics: There is no proper phase space in quantum systems - Heisenberg's uncertainty principle forbids that a point in phase space (uncertainty zero in position and momentum) exists.   
Heisenberg's uncertainty relation is a direct consequence of quantum mechanical fluctuations. A common approach to describe quantum chaos is random matrix theory and the use of energy level spacing distributions \cite{Guhr98}. A fundamental conjecture by Bohigas, Giannoni and Schmit \cite{Bohigas84} postulates that the energy level spacing distributions possesses a dominant part -  which depends on the particular system - and a subleading universal part - independent of the particular system.
The universal part gives a level spacing distribution of Wignerian type, if the corresponding classical system is fully chaotic. Such level spacing distribution can be generated also by random matrices of a certain symmetry class.  
The conjecture has not been rigorously proven yet, but has been verified and found to be valid in almost all cases. 

This approach is successful and very popular. However, it has some short comings: First, the conjecture by Bohigas et al. holds strictly only in the case of a fully chaotic system, while in nature most systems are only partly chaotic, i.e. so-called mixed systems where its classical counter part has coexistence of regular and chaotic phase space. In such cases the quantum system yields a level spacing distribution, which is neither Wignerian nor Poissonian (where the latter corresponds to a completely regular system). There is no mathematical prediction of the functional form of such distribution. However, a number of interpolations between the Poisson and Wigner distribution have been proposed \cite{Percival73,Brody81,Izrailev90,Lenz91,Berry84}. Second, we may ask: How about the comparison of the classical with the quantum system? And what is the quantitative degree of chaos? How can we answer this when the instruments used to measure chaos are quite different for both systems? 

Starting from this perspective and having in mind the goal to compare classical with quantum chaos, one may try the following strategy: 
Find a uniform description of chaotic phenomena, valid for both, classical and quantum systems. In more detail: Starting from non-linear dynamics and phase space in classical systems, one may look for a suitable quantum analogue phase space. Starting from random matrix theory and energy level spacing distributions of quantum systems, one may seek a random matrix description and a level spacing distribution of suitable dynamical objects in classical physics. 
In the following we will discuss some progress recently made in this direction.
Using some of those results, we will compare for a particular system the chaotic behavior of the quantum system with the classsical system. The numerical analysis shows that the quantum system is globally less chaotic than the classical system. We believe that such finding is not limited to the particular system. In particular, we want to understand the underlying reason for such behavior.
\begin{figure}
\begin{center}
\includegraphics[height=6cm,scale=0.15,angle=0]{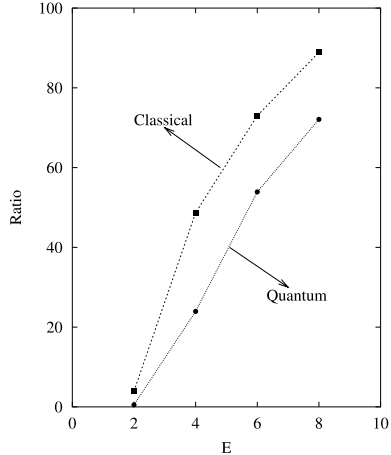}
\caption{Volume of chaotic phase space over total phase space versus energy.
Comparison of results from classical action (dotted line) with quantum action (full line).}
\label{fig:RatioPhase} 
\end{center}      
\end{figure}

\section{Cases where quantum chaos was found to be weaker}
\label{sec:Weaker}
\begin{figure}
\centering
\includegraphics[height=6cm]{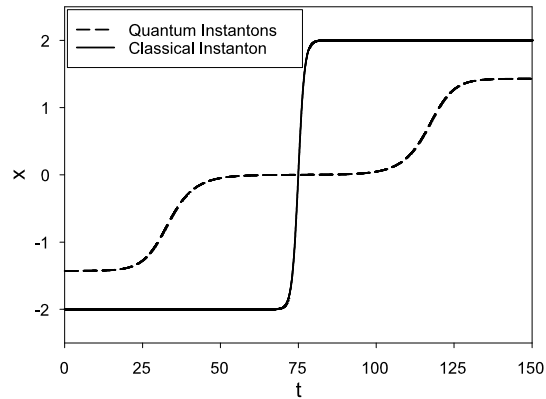}
\caption{Comparison of classical instanton with instanton from the quantum action.}
\label{fig:Instanton}      
\end{figure}

First, Casetti et al. \cite{Casetti} considered the $N$-component $\Phi^{4}$ theory in the presence of an external field and in the limit of large $N$. 
They used mean field theory and observed a strong suppression of chaos in the quantum system, due to quantum corrections causing the system the system      
to move away from a hyperbolic fixed point responsable for classical chaos.
Second, Matinyan and M\"uller \cite{Matinyan} studied another field theoretic model which is classically chaotic, namely massless scalar electrodynamics. They investigated the corresponding quantum field theory using effective field theory and loop expansion. They noticed that quantum corrections increase the threshold for chaos due to a modification of the ground state of the system.
A third example is the kicked rotor which is a classically chaotic system in 1-D. Schwengelbeck and Faisal \cite{Schwengelbeck} considered the corresponding quantum system using Bohm's interpretation of quantum mechanics to introduce trajectories and a quantum equivalent phase space. They found that the Kolmogorov-Sinai entropy goes to zero in the quantum system, i.e. it is  non-chaotic. 
This approach has been applied also to study chaos in anisotropic harmonic oscillators \cite{Parmenter}, coupled anharmonic oscillators \cite{Partovi} and the hydrogen atom in an external electromagnetic field \cite{Iacomelli}.
Finally, Caron et al. \cite{Q8} considered the anharmonic oscillator in 2-D, which is classically a mixed chaotic system. Using the concept of the quantum action functional, a quantum analogue phase space has been constructed. As a result, the phase space portrait of the quantum system was found to be slightly but globally less chaotic for all energies (see Fig.[\ref{fig:RatioPhase}]). 
\begin{figure}
\centering
\includegraphics[height=6cm]{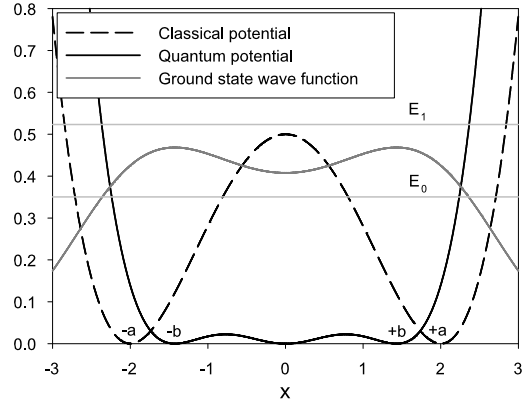}
\caption{Comparison of classical potential and quantum potential (multiplied by a scale factor 1.7331) for quartic coupling $\lambda=\frac{1}{32}$. Also shown is the ground state wave function and two lowest energy levels.}
\label{fig:TunnelPot}      
\end{figure}
Because the quantum action has been constructed from the classical action by taking into account quantum fluctuations \cite{Q1,Q2,Q3,Q4,Q5,Q6,Q7}, hence the softening of chaos in the quantum system must be due to quantum fluctuations. We suspect that such softening effect may not solely show up in chaos. Indeed, looking at a double-well potential $V(x) = \lambda \Big( x^2 - \frac{1}{8\lambda} \Big)^{2}$ in the context of tunneling, we observe that quantum effects cause the quantum potential to be much "weaker" than the classical potential (see Fig.[\ref{fig:TunnelPot}]), i.e. the potential wells are less pronounced and the potential barrier is much lower for the quantum potential (note that the quantum potential has a triple-well shape).  
The shape of the potential for tunneling translates into the shape of the instantons. Thus it comes as no surprise that the instanton of the quantum action (actually a double-instanton) is softer than the classical instanton (see Fig.[\ref{fig:Instanton}]).

\section{Uniform description of chaos}
\label{sec:Uniform}
Let us recall that the conjecture by Bohigas et al. is about random matrix theory and energy level spacing distributions to describe chaos in quantum systems, while chaos in classical systems is conventionally described in terms of phase space. That is, the tools to describe chaos in classical and quantum physics are different. For the purpose to better understand the physical content of this conjecture or eventually to find a mathematical proof, it is highly desirable to use the same language, respectively, tools in quantum physics as in classical physics. This can be viewed in two ways: Either one adopts the point of view that chaos should be analyzed in terms of phase space. Then a uniform description is achieved by use of classical phase space in classical physics and the quantum analogue phase space in quantum systems. How about the point of view that chaos should be analyzed in terms of random matrix theory and level spacing distributions? In the following we propose how to construct such a level 
spacing distribution, which will play the same role in classical physics as the energy level spacing distribution in the quantum system.

One may immediately object that  the energy level spacing distribution in quantum physics is due to the fact that one considers a system of bound states (with a "confining" potential), which due to the rules of quantum mechanics gives a discrete spectrum. The discreteness of the spectrum is a quantum effect, i.e., a physical effect.
Having in mind to construct a counterpart in classical physics, one may object that classical physics is continuous. There is no discreteness inherent in classical physics. If any discreteness occurs it will be due to some (mathematical) approximation. Hence, how can a level spacing distribution derived from a classical function be discrete and physical?
First we want to propose a function and show how to construct a meaningful level spacing distribution. Afterwards, we will try to answer the last question.

We would like to emphasize two properties: First, the level spacing distribution of the quantum system - in the case of a classically fully chaotic system - 
corresponds to a Wigner distribution. The type of Wigner distribution is determined by the symmetry of the Hamiltonian. E.g., there is the Gaussian orthogonal ensemble (GOE), the unitary ensemble and the symplectic ensemble.
Thus, we need a function which carries the same symmetries. A function convenient for this purpose is the classical action.
Another property to be emphasized is locality. Let us recall that the level spacing distribution, e.g. the GOE distribution, is invariant under orthogonal transformations. Because an orthogonal transformation maps any orthogonal basis of states onto another orthogonal basis, the level spacing distribution is essentially independent of the particular choice of the basis.
That means, we may choose a basis that is quasi-local, i.e., built from square integrable functions which are identically zero everywhere except in a small interval where they are non zero and constant. Those box functions are almost local. The advantage of locality is the fact that symmetries of the Hamiltonian often have to do with transformations in position space, and locality facilitates the analysis of such symmetries. Why do we need to bother about symmetries? Because the energy level spacing distribution is meaningful only (and gives the Wigner distribution for a chaotic system, respectively a Poissonian distribution for an integrable system) if the energy levels have all the same quantum numbers (exept for the quantum number of energy). For example, for the spectrum of the hydrogen atom, one should take the bound states all with the same angular momentum quantum numbers, e.g., $l=m=0$.

\subsection{Action matrix}
\label{sec:Action}
We consider a system defined by a Langrange function $L(q,\dot{q},t)$. Let $S$ denote the corresponding action,
\begin{equation}
S[q(t)] = \int_{0}^{T} dt ~ L(q,\dot{q},t) ~ .
\end{equation}
Let $q_{traj}$ denote the trajectory, i.e. the solution of the Euler-Lagrange equation of motion. Such trajectory is a function, which makes the action functional stationary. Each trajectory is specified by indicating the initial and final boundary points, i.e., $q(t=0) = q_{in}$ and $q(t=T) = q_{fi}$. 
We assume that those boundary points are taken from some finite set of nodes, 
$q_{in},q_{fi} \in \{q_1,\dots,q_N\}$. Thus with each pair of boundary points, $(q_{k},q_{l})$  we associate a trajectory $q^{traj}_{kl}(t)$.
Then we introduce an action matrix $\Sigma$, where the matrix element $\Sigma_{kl}$ corresponds to the value of the action $S$ evaluated along the trajectory $q^{traj}_{kl}(t)$,
\begin{equation}
\Sigma_{kl} = S[q^{traj}_{kl}(t)] ~ .
\end{equation}
All matrix elements $\Sigma_{kl}$ are real. They are also symmetric, $\Sigma_{kl}=\Sigma_{lk}$. Thus the action matrix $\Sigma$ is a Hermitian 
$N \times N$ matrix. Consequently, $\Sigma$ has a discrete spectrum of action eigenvalues, which are all real,  
\begin{equation}
\sigma(\Sigma) = \{\sigma_1,\dots,\sigma_N\} ~ .
\end{equation}

\subsection{Symmetry}
\label{sec:Symmetry}
For the purpose to compute a level spacing distribution from the action eigenvalues, one must first address the issue of symmetry. Let us consider for example the harmonic oscillator in 2-D. We may take the coordinates $q_{k}$ to be located on the nodes of a regular grid (reaching from $-\Lambda$ to $+\Lambda$ on the $x$ and $y$ axis), with a spacing $\Delta x = \Delta y = a = const$. For example, the action of the harmonic oscillator, with the classical trajectory going from $\vec{x}_{a}$ to $\vec{x}_{b}$ in time $T$ is given by
\begin{equation}
\Sigma_{a,b} = \frac{m \omega}{2 \sin(\omega T)} \left[ (x^{2}_{a} + x^{2}_{b}) \cos(\omega T) - 2 \vec{x}_{a} \cdot \vec{x}_{b} \right] ~ .
\end{equation}
This function is invariant under rotations. When choosing the coordinates $q_{k}$ to be located on the regular grid, the continuous symmetry of rotation will become a discrete symmetry of finite rotations (a group). We have to find the irreducible representations of such group and sort the action eigenvalues according to those irreducible representations (this can be done by inspecting the properties of the corresponding eigenvector). Then one has to select a particular representation and retain a subset of eigenvalues in that representation. The action level spacing distribution can then be obtained from the action eigenvalues in such subset.

This procedure is feasible. However, it has two disadvantages. First, finding the irreducible representations and classifying the eigenvectors accordingly is laborious. More importantly, the fact that one has to work with a subset of eigenvuales only and not the whole ensemble of eigenvalues means a drastic reduction of the size of the statistical ensemble.
In other words, the statistics of the resulting level spacing distribution will
deteriorate.

For those reasons it would be highly desirable to avoid the above strategy. This is indeed possible by using the following trick. One can camouflage the symmetry by choosing the coordinates off the nodes of the regular grid. That means, for example to define new coordinates as follows,
\begin{equation}
q^{deform}_{k} = q_{k} + \epsilon_{k} ~ ,
\end{equation}
where $\epsilon_{k}$ denotes a randomly chosen small deformation (in angle and length). In this way the nodes are irregularly distributed. Consequently, the discrete symmetry of the action matrix $\Sigma_{kl}$ disappears when replacing it by the "deformed" action matrix,
\begin{eqnarray}
\Sigma^{deform}_{kl} 
&=& \mbox{action evaluated along the classical trajectory} 
\nonumber \\
& & \mbox{from node} ~ q^{deform}_{k} ~ \mbox{to node} ~ q^{deform}_{l} ~ .
\end{eqnarray}
In doing so we avoid a laborious symmetry analysis and secondly will have a better statistics!

\subsection{Action level spacing distribution}
\label{sec:ActionLSD}
Now we want to construct a level spacing distribution of action levels. 
We proceed in analogy to random matrix theory and the method of constructing an energy level spacing distribution. For an overview on how to compute level spacing distributions see Ref.\cite{Haake01}. 
One has to separate the dominant system dependent part from the subleading universal part which describes the properly normalized fluctuations.  
Because we are here interested only in the fluctuation part, we suppress the leading part. One should note that this means to discard all physical information which depends on the particular system. For example thermodynamical functions can not be computed from the subleading fluctuating part.
The strategy to obtain the subleading part is called unfolding. One constructs a fit to the original spectrum and multiplies the spectrum such that on average the mean spacing distribution becomes unity. Also the integrated level spacing distribution will be normalized to unity. 

We have applied this to simple integrable systems in 1-D. For integrable systems one would expext a Poissonian distribution for the action level spacing distribution. Preliminary results are compatible with a Poissonian distribution. 
The following remarks are in order. The first numerical results for integrable systems have to be repeated with precision and analyzed carefully. Second, one wants to see what happens in chaotic systems. Possibly such strategy applied to a fully chaotic system will result in a Wignerian action level spacing distribution. Finding an answer will be computationally much more involved, simply because the action functions for the integrable systems considered above are analytically known, while for a chaotic system (i.e. non-integrable) this needs to be calculated numerically. Moreover, the numerical precision required needs to be sufficient to resolve small fluctuations. In the statistical sense, one is interested in a sample of large size. But that means that after unfolding the fluctuations will become small and hence require a high numerical precision for its resolution. Presently, numerical studies of such question are under way.  

Let us get back to the question posed at the beginning of Sect.\ref{sec:Uniform}:
How can a level spacing distribution derived from a classical function be discrete and physical? In our opinion, the answer lies in the fact that the level spacing distribution is universal, that is, it does not depend, for example, on the parameters of the discrete grid. Different grids give the same result. This has been verified numerically. It also should not depend on the deformation $\epsilon_{k}$ (as long it is not too close to zero and the discrete symmetry is restored). Also this has been verified numerically and found to be satisfied. 
Thus one could in principle go with the volume of the lattice $V=(2\Lambda)^D$ to infinity and with the lattice spacing $\Delta x = \Delta y$ to zero. The result should not change, but one would have reached the continuum limit. The discreteness would then disappear.

\section{Renormalisation flow of parameters of the quantum action}
\label{sec:Renormalisation}
In Sect.\ref{sec:Weaker} we have seen examples for the observation that quantum chaos seems to be weaker than chaos in the corresponding classical system. 
Of course it would be interesting to explore a much wider class of systems in order to see if such observation holds more generally. Here we want to  
pick one of the above examples, namely the the chaotic anharmonic oscillator in 2-D and try to understand why quantum chaos is weaker than classical chaos. 
The classical action is given 
\begin{eqnarray}
S &=& \int_{0}^{T} dt \frac{1}{2m} (\dot{x}^2 + \dot{y}^2) - V(x,y) ~ ,
\nonumber \\
V &=& \frac{m \omega^2}{2} (x^2 + y^2) + \lambda x^2y^2 
= v_0 + v_2 (x^2 + y^2) + v_{22} x^2y^2 ~ . 
\end{eqnarray}
For $\lambda=0$ the system is reduced to the standard harmonic oscillator, which is integrable. The chaoticity is introduced and controlled by the parameter $\lambda$. Thus small $\lambda$ causes mild chaos, while large $\lambda$ makes the system strongly chaotic. 
The quantum action has been postulated to be of the functional form like the classical action, i.e. the kinetic term of the quantum action may differ in the value of the mass, and the potential of the quantum action should also be local, depend only on coordinates, but may have a different functional form.
Here let us consider an ansatz of the following form
\begin{eqnarray}
\tilde{S} &=& \int_{0}^{T} dt \frac{1}{2\tilde{m}} (\dot{x}^2 + \dot{y}^2) - \tilde{V}(x,y) ~ ,
\nonumber \\
\tilde{V} &=& \tilde{v}_0 + \tilde{v}_2 (x^2 + y^2) + \tilde{v}_{22} x^2y^2 + 
\mbox{higher order polynomials} ~ .
\end{eqnarray}

As a quantitative measure of the strength of chaos we take the strength of the parameters of the action, in particular, the parameter $\lambda$.   
The parameters of the quantum action can be interpreted as a "renormalisation effect" of the parameters of the classical action. The calculation of those parameters has to be done numerically, following the definition of the quantum action to be a functional which fits the transition amplitudes \cite{Q1}. 
However, in a certain limit, the quantum action is known to be an exact representation of the transition amplitudes and moreover the action is related via differential equations to the classical action \cite{Q4}. This limiting case is using imaginary time and let time go to infinity (Feynman-Kac limit). 
Because we want to obtain an analytical result, we will use perturbation theory. This means that we consider the regime of small $\lambda = v_{22}$.

In order to simplify the matter, let us start by considering the system in 1-D.
Thus we have the potential
\begin{equation}
V(x) = \frac{1}{2} m \omega^{2} x^{2} + \lambda x^{4} 
\equiv v_2 x^2 + v_4 x^4 ~ .
\label{eq:DefV}
\end{equation}
We assume $\lambda$ to be small,
\begin{equation}
\frac{\lambda \Lambda_{sc}}{v_{2}} << 1 ~ ,
\end{equation}
where $\Lambda_{sc}$ introduces a physical length scale, e.g. the analogue of the Bohr radius. According to Ref.\cite{Q5} the following relation between the classical and the quantum potential holds,
\begin{equation}
2 m(V(x) - E_{gr}) =  
2 \tilde{m}(\tilde{V}(x) - \tilde{v}_{0}) 
- \frac{\hbar}{2} \frac{ \frac{d}{dx} 2 \tilde{m} (\tilde{V}(x) - \tilde{v}_{0})}
{ \sqrt{2 \tilde{m}( \tilde{V}(x) - \tilde{v}_{0} ) } } ~ \mbox{sgn}(x) ~ .
\label{eq:TransformLaw}
\end{equation}
We define the functions
\begin{eqnarray}
W(x) &=& 2m(V(x) - E_{gr}) ~ ,
\\
U(x) &=& 2\tilde{m}(\tilde{V}(x) - \tilde{v}_{0}) ~ ,
\label{eq:DefW+U}
\end{eqnarray}
where $E_{gr}$ denotes the ground state energy of the lowest eigen state of the quantum mechanical system (obtained from the Schr\"odinger equation). Due to Eq.(\ref{eq:DefV}), the function $W(x)$ must have the form
\begin{equation}
W(x) = w_{0} + w_{2} x^{2} + w_{4} x^{4} ~ .
\label{eq:AnsatzW}
\end{equation}
This function is symmetric with respect to parity.
Then the function $U(x)$ representing the quantum potential, will be parity symmetric also. We make an Ansatz of the form
\begin{equation}
U(x) = u_{0} + u_{2} x^{2} + u_{4} x^{4} + u_{6} x^{6} + \cdots 
\label{eq:AnsatzU}
\end{equation}
The assumption that the expansion parameter $\lambda$ is small is now expressed by
\begin{equation}
w_{4} \equiv w^{(0)}_{4} \epsilon ~, ~~~ \epsilon << 1 
\label{eq:DefW04}
\end{equation}
Now using Eqs.(\ref{eq:AnsatzW},\ref{eq:AnsatzU}) in combination with Eq.(\ref{eq:TransformLaw}), we obtain
\begin{eqnarray}
w_{0} + w_{2} x^{2} + w_{4} x^{4} 
&=&  u_{2} x^{2} + u_{4} x^{4} + u_{6} x^{6} + \cdots
\nonumber \\
&-& \frac{\hbar}{2} 
\frac{ 2 u_{2} + 4 u_{4} x^{2} + 6 u_{6} x^{4} + \cdots }
{ \sqrt{ u_{2} + u_{4} x^{2} + u_{6} x^{4} + \cdots } } ~~~ 
\mbox{for} ~ x > 0 ~ . 
\label{eq:TransformWU}
\end{eqnarray}
The smallness of $w_4$ implies that the terms of fourth order and higher in $x$ occuring in the function $U(x)$ are small compared to the second order, i.e.
\begin{equation}
u_{4} x^{2} + u_{6} x^{4} + \cdots << u_{2} ~ .
\end{equation}
Now doing a Taylor expansion in the small terms 
$u_{4} x^{2} + u_{6} x^{4} + \cdots$ 
allows to express the r.h.s. of Eq.(\ref{eq:TransformWU}) 
as a polynomial in $x$. Then comparing terms in $x$ order by order, we find the following relations
\begin{eqnarray}
w_{0} &=& - \frac{ \hbar }{ 4 \sqrt{u_{2}}^{3} } ~ 4 u_{2}^{2} ~ ,
\nonumber \\
w_{2} &=& u_{2} - \frac{ \hbar }{ 4 \sqrt{u_{2}}^{3} } ~ 6 u_{2} u_{4} ~ ,
\nonumber \\
w_{4} &=& u_{4} - \frac{ \hbar }{ 4 \sqrt{u_{2}}^{3} } (10 u_{2} u_{6} - 4 u_{4}^{2}) ~ .
\label{eq:RelationCoeff}
\end{eqnarray}
Now we try to find the parameters $u_{2}$ and $u_{4}$ as solution of those equations. The first equation gives
\begin{equation}
u_{2} = (w_{0}/\hbar)^{2} ~ .
\end{equation}
On the other hand we have, due to Eq.(\ref{eq:DefW+U}), 
\begin{equation}  
w_{0} = - 2 m E_{gr} ~ .
\label{eq:w0}
\end{equation}
Due to the anharmonic perturbation the ground state energy is different from the ground state energy of the harmonic oscillator. However, because of the smallness of the perturbation, we can express the groundstate energy $E_{gr}$ using perturbation theory as a power series in $\epsilon$,
\begin{equation}
E_{gr} = E^{(0)} + \epsilon E^{(1)} + \epsilon^{2} E^{(2)} + \cdots ,
\label{eq:GroundStExp}  
\end{equation}
where $E^{(0)}= E^{osc}_{gr}$. Thus from Eqs.(\ref{eq:w0},\ref{eq:GroundStExp}) we obtain
\begin{equation}
\sqrt{u_{2}} = m \omega + \epsilon \frac{2 m E^{(1)} }{\hbar} 
+ O(\epsilon^{2}) ~ ,
\end{equation}
or
\begin{eqnarray}
u_{2} &=& m^{2} \omega^{2} + \epsilon \frac{4 m^{2} E^{(1)} }{\hbar} 
+ O(\epsilon^{2})
\nonumber \\ 
&=& w_{2} [ 1 +  \frac{2 E^{(1)}}{E^{(0)}} \epsilon + O(\epsilon^{2}) ] ~ .
\label{eq:SolutionU2}
\end{eqnarray} 
Next let us consider Eq.(\ref{eq:RelationCoeff}b). 
We obtain
\begin{equation}
u_{4} = - \frac{2}{3 \hbar} \sqrt{u_{2}} (w_{2} - u_{2}) ~ .
\end{equation}
Recalling $w_{2} = m^{2} \omega^{2}$ and Eq.(\ref{eq:SolutionU2}), we find
\begin{equation}
u_{4} = - \frac{2}{3} \frac{m^{3} \omega^{3}}{\hbar} 
\frac{E^{(1)}}{\hbar \omega} \epsilon + O(\epsilon^{2}) ~ .
\label{eq:SolutionU4}
\end{equation}
In Eqs.(\ref{eq:SolutionU2},\ref{eq:SolutionU4}) we have expressed 
the parameters of the quantum action in terms of the parameters of the classical action. However, it remains to compute the energy $E^{(1)}$. Again we use stationary perturbation theory.
The Hamiltonian is given by, taking into account Eqs.(\ref{eq:AnsatzW},\ref{eq:DefW04}),
\begin{eqnarray}
H &=& H^{(0)} + \epsilon H^{(1)}
\nonumber \\
H^{(0)} &=& \frac{p^{2}}{2 m} + \frac{1}{2} m \omega^{2} x^{2} 
\nonumber \\
H^{(1)} &=& \frac{w^{(0)}_{4}}{2m} x^{4} ~ .
\end{eqnarray}
To first order of perturbation theory in $\epsilon$ the energy $E^{(1)}$ is given by
\begin{equation}
E^{(1)} = \langle \psi^{osc}_{gr}| \frac{w^{(0)}_{4}}{2m} x^{4} | \psi^{osc}_{gr} \rangle ~ .
\end{equation}
which yields the result
\begin{equation}
E^{(1)} = \frac{ 3 w^{(0)}_{4} \hbar^{2} }{ 8 m^{3} \omega^{2} } ~ .
\label{eq:ResultE1}
\end{equation}
Substituting this result into Eqs.(\ref{eq:SolutionU2},\ref{eq:SolutionU4}) we finally obtain 
\begin{equation}
u_{2} = w_{2} \left[ 1 + (\frac{3 \hbar}{2 m^{3} \omega^{3}} ) w_{4} + O(\epsilon^{2}) \right] ~ ,
\label{eq:FinalU2}
\end{equation}
and
\begin{equation}
u_{4} = - \frac{1}{4} w_{4} + O(\epsilon^{2}) ~ .
\label{eq:FinalU4}
\end{equation}

\section{Interpretation}
\label{sec:Interpretation}
Let us see what happens when we keep the classical parameters fixed, exept for $w_{4}$, i.e., we keep $w^{(0)}_{4}$ fixed and vary $\epsilon$. Note that 
$w_{2} > 0$ and $w_{4} > 0$. We also have $w^{(0)}_{4} > 0$ and $\epsilon > 0$.
Now we want to study what happens when $\epsilon \to 0$.
Eq.(\ref{eq:FinalU2}) yields
\begin{eqnarray}
&& u_{2} > w_{2} 
\nonumber \\
&& u_{2} \longrightarrow_{\epsilon \to 0} w_{2} ~ .
\end{eqnarray}
Likewise, Eq.(\ref{eq:FinalU4}) yields
\begin{eqnarray}
&& u_{4} < w_{4} 
\nonumber \\
&& u_{4} \longrightarrow_{\epsilon \to 0} w_{4} 
\longrightarrow_{\epsilon \to 0} 0 ~ .
\end{eqnarray}
In other words, in the limit $\epsilon \to 0$ the classical potential approaches the potential of the harmonic oscillator. The potential of the quantum action asymptotically approaches the classical potential, hence also the harmonic oscillator potential. That is, the renormalisation group flow 
of the parameters $u_{2}(\epsilon),u_{4}(\epsilon)$ goes to a Gaussian fixed point. Second, for any value of $\epsilon$ the value the quadratic term of the potential is larger for the quantum potential than for the classical potential.
Third, for any value of $\epsilon$ the value the quartic term of the potential is smaller for the quantum potential than for the classical potential.
Recall that the quadratic term is the term, which, if it would stand alone, would make the system integrable. On the other hand, the quartic term is the term which drives the system away from integrability (and introduces chaos in 2-D). Thus we find that quantum fluctuations, which are the cause for the differences $\Delta_{2} = w_{2} - u_{2}$ and $\Delta_{4} = w_{4} - u_{4}$ to be non-zero, have the tendency to drive the quantum system closer to the regime of integrability. 

Now the above perturbative calculations were performed in 1-D. Chaos in time-independent Hamilton systems exists only for $D \ge 2$. A similar, but more tedious calculation can be performed in 2-D. It confirms the above result that   quantum fluctuations drive the system closer to the regime of integrability and away from the regime of chaos.

Of course such perturbative calculations are meaningful only in a neighborhood of the Gaussian fixed point. It would be desirable to extend the calculations to a larger regime. However going to higher order of perturbation theory would make those calculation much more tedious. Nevertheless the perturbative result gives some insight into the dynamical consequences of quantum fluctuations.
\\

\noindent Acknowledgment. H.K. has been supported by NSERC Canada.
\\

\end{document}